\documentclass[a4paper]{article}

\usepackage{INTERSPEECH2022}
\usepackage{algorithm}
\usepackage{algpseudocode}
\usepackage{multirow}
\usepackage{array}
\usepackage{cite}

\usepackage{xcolor}

\title{Transport-Oriented Feature Aggregation for Speaker Embedding Learning}
\name{Yusheng Tian, Jingyu Li, Tan Lee}
\address{
  Department of Electronic Engineering, The Chinese University of Hong Kong, Hong Kong, China
  }
\email{\{ystian0617, lijingyu0125\}@link.cuhk.edu.hk, tanlee@ee.cuhk.edu.hk}

\begin{document}

\maketitle
\begin{abstract}
Pooling is needed to aggregate frame-level features into utterance-level representations for speaker modeling. Given the success of statistics-based pooling methods, we hypothesize that speaker characteristics are well represented in the statistical distribution over the pre-aggregation layer's output, and propose to use transport-oriented feature aggregation for deriving speaker embeddings. The aggregated representation encodes the geometric structure of the underlying feature distribution, which is expected to contain valuable speaker-specific information that may not be represented by the commonly used statistical measures like mean and variance. The original transport-oriented feature aggregation is also extended to a weighted-frame version to incorporate the attention mechanism. Experiments on speaker verification with the Voxceleb dataset show improvement over statistics pooling and its attentive variant.

\end{abstract}

\section{Introduction}

Speaker embedding learning, i.e., modeling speaker identity with a compact vector representation, is of interest in many tasks of speech and language processing. Examples are speaker verification \cite{gmmubm}, speaker diarization \cite{diarization}, and voice conversion/cloning \cite{voice_conversion, voice_cloning}. In the past decade, state-of-the-art methods of speaker embedding learning have shifted from statistical model-based \cite{gmmubm, ivector} to deep neural network (DNN)-based \cite{dvector, dvector2,xvectorbase,voxceleb2,ecapa}. The DNN-based embedding extraction models are typically trained to minimize speaker classification loss, and they make use of a pooling layer to aggregate the variable-length sequence of features into a fixed-size representation. The most common pooling method is known as statistics pooling \cite{xvectorbase, xvector}, which computes the mean and standard deviation of frame-level features. Channel-wise correlation-based pooling has also been shown effective in \cite{correlation}.

Inspired by the success of statistics-based pooling methods, we hypothesize that speaker characteristics are well represented in the statistical distribution over the pre-aggregation layer's output, and that speaker discrimination can be achieved based on measured distance between the respective distributions. Although intra-speaker factors, such as the change of speaking style, might also cause variation in feature distributions, supervised training with speaker labels is expected to suppress their effect. The above hypothesis can also be interpreted as drawing an analogy between voice characteristics and image style, which was shown to be strongly related to feature distributions in computer vision research \cite{demystify, relaxedOT}.

Following this hypothesis, we propose the approach of transport-oriented feature aggregation \cite{LOT, graph, OTKE} to speaker embedding learning. Given a set of feature samples drawn from certain distribution, transport-oriented aggregation produces an isometric linear Wasserstein embedding. This embedding encodes the geometric structure of the underlying feature distribution, which is expected to contain valuable speaker-specific information that may not be represented by the commonly used statistical measures like mean and variance. Furthermore, the Euclidean distance of two embedded feature distributions is an approximation of their Wasserstein (a.k.a optimal transport \cite{villani2009optimal}, or earth mover's \cite{rubner1998metric}) distance. This allows us to use the cosine similarity to measure the distance between two feature distributions with their $l_2$-normalized embeddings. 

To verify the advantage of transport-oriented feature aggregation over statistics pooling, a toy example is first studied on the task of classifying distributions of 1D features. The example gives us insights on how the transport-based model behaves in general and how to apply it. We then apply transport-oriented feature aggregation to speaker verification, to evaluate the quality of the resulting speaker embeddings. The original transport-oriented feature aggregation is also extended to a weighted-frame version to incorporate the attention mechanism. Experiments on speaker verification with the Voxceleb dataset \cite{voxceleb1, voxceleb2} show improvement over statistics pooling and its attentive variant.

The contributions of this paper are summarized as follows. First, a novel feature aggregation approach to deriving speaker embeddings is presented and investigated. The underlying idea is to distinguish speech from different speakers by measuring the optimal transport distance between the corresponding feature distributions. Second, the original transport-oriented feature aggregation is extended to an attention-weighted version, which leads to further improvement on the model performance.

\section{Related work}
\subsection{Pooling in speaker embedding networks}
The most common choice of pooling is statistics pooling introduced as in the seminal work on x-vector \cite{xvectorbase, xvector}. Pooling over frame-level representations is implemented by computing and concatenating the utterance-wide mean and standard deviation. Subsequent work built upon statistics pooling explored improvement mainly by incorporating the attention mechanism \cite{att1, att2, att-jingyu}, by which trainable importance weights are assigned to individual frames. Another direction of research is about replacing the mean and standard deviation by other statistics. Wang et al. \cite{covariance} investigated pooling with high-order statistics, e.g., covariance and $l_p$-norm. The experimental results showed that using high-order statistics yielded consistently better performance than simple average pooling. In the recent work \cite{correlation} by Stafylakis et al., channel-wise correlations were proposed as the statistical measures for pooling and outperformed statistics pooling under certain configurations. The proposed transport-oriented approach differs from these statistics-based pooling methods in that speaker characteristics are represented by the entire feature distribution, rather than a few selected statistical parameters.

Another strand of work on pooling applies residual encoding to speaker modeling, which learns a multi-modal reference and uses the accumulated residuals (differences between frame-level features and the reference centers) as speaker representations. Examples include Net/GhostVLAD \cite{netvladaudio} and LDE \cite{lde}. They are more commonly seen in 2D-CNN based speaker embedding networks as these methods originated from computer vision applications like image retrieval \cite{netvlad} and texture classification \cite{deepten}. Broadly speaking, these methods are in the same spirit as the classical i-vector \cite{ivector}, which assumes Gaussian Mixture distribution of features. The proposed method in this paper is also designed for 2D-CNN based networks, and can be viewed as a type of residual encoding. However, no specific assumption is required on the type of feature distribution.

\subsection{Wasserstein distance}
Comparing probability distributions is a fundamental problem in machine learning. The Wasserstein distance has become a popular metric. For example, in the Wasserstein GAN framework \cite{wgan}  the discriminator computes the Wasserstein distance between the generated and the real data distributions. In \cite{wmd}, Kusner et al. used the Wasserstein distance of word distributions to measure dissimilarity between text documents.

\subsection{Linear Wasserstein embedding}
Pooling is needed in order to aggregate a set of features into a fixed-size embedding for downstream tasks.  It is desirable that the statistics of features to be aggregated are optimally preserved in the embedding. The linear Wasserstein embedding \cite{LOT} can meet the need: the Wasserstein distance between two feature distributions can be estimated by the $l_2$ distance of their pooled representations. The key steps are to determine a reference distribution and compute the optimal transport plan between the input and this reference. Originally the reference distribution is learned offline by k-means clustering \cite{LOT, graph}. Very recently, Milan et al. \cite{OTKE} proposed to learn the reference distribution in conjunction with the downstream task. We adopt this form of linear Wasserstein embedding in this study.

\section{Transport-oriented feature aggregation} \label{approach}
In this section, a brief review of the theoretical foundation of transport-oriented feature aggregation is first given. Then we describe how it is applied to speaker embedding learning.

\subsection{Theoretical foundation}
Suppose we are interested in a $d$-dimensional feature space. Consider a set $\mathbf{x}$ containing $N_x$ feature samples, i.e., $\mathbf{x}=\{\mathbf{x}^{(1)}, \mathbf{x}^{(2)}, ..., \mathbf{x}^{(N_x)}\}$ with $\mathbf{x}^{(i)}\in\mathbb{R}^d$. The underlying distribution of these features is approximated by the discrete probability measure $\mu_x=\sum_{i=1}^{N_x} a_i\delta_{\mathbf{x}^{(i)}}$, where $\delta_{\mathbf{x}^{(i)}}$ is the Dirac function at location $\mathbf{x}^{(i)}$, and $a_i$ the intensity. For equally weighted observations, we have $a_i=1/N_x$. 

Let $\mathbf{z}=\{\mathbf{z}^{(1)}, \mathbf{z}^{(2)}, ..., \mathbf{z}^{(N_z)}\}$ be a reference set defined on the same feature space. It has $N_z$ samples drawn from a reference distribution, representing the discrete probability measure $\mu_z=\sum_{j=1}^{N_z} b_j\delta_{\mathbf{z}^{(j)}}$. The Wasserstein distance between $\mu_x$ and $\mu_z$ is obtained by solving the following optimal transport problem of moving the mass from $\mathbf{x}$ to $\mathbf{z}$:
\begin{equation} \label{eq:wasserstein_distance}
    \mathcal{W}_2^2(\mu_x,\mu_z)=\min_{\mathbf{P}}\sum_{i=1}^{N_x}\sum_{j=1}^{N_z}\mathbf{P}_{ij}\Vert \mathbf{x}^{(i)}-\mathbf{z}^{(j)}\Vert^2 .
\end{equation}
The search space of the transport plan $\mathbf{P}(\mathbf{x}, \mathbf{z})\in\mathbb{R}^{N_x\times N_z}$ is constrained by $\sum_{j=1}^{N_z}\mathbf{P}_{ij}=a_i, \sum_{i=1}^{N_x}\mathbf{P}_{ij}=b_j$. The linear Wasserstein embedding of $\mathbf{x}$ is given by the optimal transport plan, denoted as $\mathbf{P}^*(\cdot)$, between $\mathbf{x}$ and the reference $\mathbf{z}$:
\begin{equation}
    \phi(\mathbf{x})=[\phi(\mathbf{x})^{(1)},\phi(\mathbf{x})^{(2)}, \ldots, \phi(\mathbf{x})^{(N_z)} ]\in\mathbb{R}^{d\times N_z},
\end{equation}
where $\phi(\mathbf{x})^{(j)}=\left(\sum_{i=1}^{N_x}\mathbf{P}^*(\mathbf{x},\mathbf{z})_{ij}\mathbf{x}^{(i)}\right)-\mathbf{z}^{(j)}\in\mathbb{R}^d$. Note that $\phi(\mathbf{x})$ has the fixed size $d\times N_z$ and that it is permutation-invariant w.r.t the elements in $\mathbf{x}$ given the reference.

Our motivation of applying this transport-oriented feature aggregation $\phi(\cdot)$ is from its nice property of preserving statistics of the input set: for any two sets $\mathbf{x}$ and $\mathbf{y}$, the Wasserstein distance between their sample distributions can be approximated by the Euclidean distance between their embeddings, i.e. $\mathcal{W}_2^2(\mu_x,\mu_y)\approx\Vert\phi(\mathbf{x})-\phi(\mathbf{y})\Vert^2$. This property allows us to measure the distance between two feature distributions by using their aggregated embeddings. The distance is used as the metric for speaker discrimination. Readers are referred to \cite{LOT, graph} for detailed derivation of $\phi(\cdot)$ and the proof of its properties.

The embedding function $\phi(\cdot)$ involves solving the optimal transport problem defined in equation (\ref{eq:wasserstein_distance}), which is costly in computation. In \cite{sinkhornOT}, Cuturi proposed to smooth the optimal transport problem using an entropy term:
\begin{equation}
    \mathcal{W}_\epsilon^2(\mu_x,\mu_z)=\min_{\mathbf{P}} \sum_{i=1}^{N_x}\sum_{j=1}^{N_z}\mathbf{P}_{ij}\Vert \mathbf{x}^{(i)}-\mathbf{z}^{(j)}\Vert^2-\epsilon H(\mathbf{P}),
\end{equation}
where $H(\mathbf{P})=-\sum_{ij}\mathbf{P}_{ij}\log( \mathbf{P}_{ij})$ is the entropy function and the value of parameter $\epsilon$ is typically in the range of 0.1-2.0. This formulation enables fast computation of the optimal transport plan using the Sinkhorn-Knopp algorithm \cite{sinkhorn1967diagonal} and the algorithm is differentiable. Milan et al. \cite{OTKE} adopted this entropic regularized optimal transport to learn the reference set with backpropagation in an end-to-end manner. We follow this formulation and the computation process of transport-oriented feature aggregation is summarized in Algorithm 1.

\begin{algorithm}
\caption{Transport-oriented feature aggregation}\label{alg:cap}
\begin{algorithmic}[1]
\Require Input set $\mathbf{x}=\{\mathbf{x}^{(i)}\}_{i=1}^{N_x}$
\Require Reference set $\mathbf{z}=\{\mathbf{z}^{(j)}\}_{j=1}^{N_z}$
\Require Intensity $\mathbf{a}\in\mathbb{R}^{N_x\times 1}$, $\mathbf{b}\in\mathbb{R}^{N_z\times 1}$
\Require Parameter $\epsilon$
\State Compute the distance matrix $\mathbf{C}\in\mathbb{R}^{N_x\times N_z}$, where $\mathbf{C}_{ij}=\Vert \mathbf{x}^{(i)} - \mathbf{z}^{j}\Vert^2$
\State $\mathbf{K} \gets \exp{(-\mathbf{C}/\epsilon)} \in \mathbb{R}^{N_x\times N_z}$
\State $\mathbf{u} \gets \mathbf{1}/N_x \in \mathbb{R}^{N_x\times 1}$
\State $\mathbf{v} \gets \mathbf{1}/N_z \in \mathbb{R}^{N_z\times 1}$
\While{not done}
\State $\mathbf{v} \gets \mathbf{b} \oslash (\mathbf{K}^{T}\mathbf{u})$ \Comment {$\oslash$ denotes element-wise division}
\State $\mathbf{u} \gets \mathbf{a} \oslash (\mathbf{K} \mathbf{v})$
\EndWhile
\State $\mathbf{P}^*=\text{diag}(\mathbf{u})\mathbf{K}\text{diag}(\mathbf{v}) \in \mathbb{R}^{N_x\times N_z}$
\State\Return $\phi(\mathbf{x})=\mathbf{X}\mathbf{P}^* - \mathbf{Z} \in \mathbb{R}^{d\times N_z}$ 
\Comment $\mathbf{X}$ and $\mathbf{Z}$ are the matrix representation for $\mathbf{x}$ and $\mathbf{z}$ respectively.
\end{algorithmic}
\end{algorithm}
\vspace{-1em}

\subsection{Tailoring for speaker embedding learning}
For speaker embedding learning, the previously defined set $\mathbf{x}$ is considered as the container of extracted features and each element of $\mathbf{x}$ corresponds to one feature at certain time frame. However, extra care must be taken to deal with speech data.

\begin{figure}[t]
  \centering
  \includegraphics[width=1.0\linewidth]{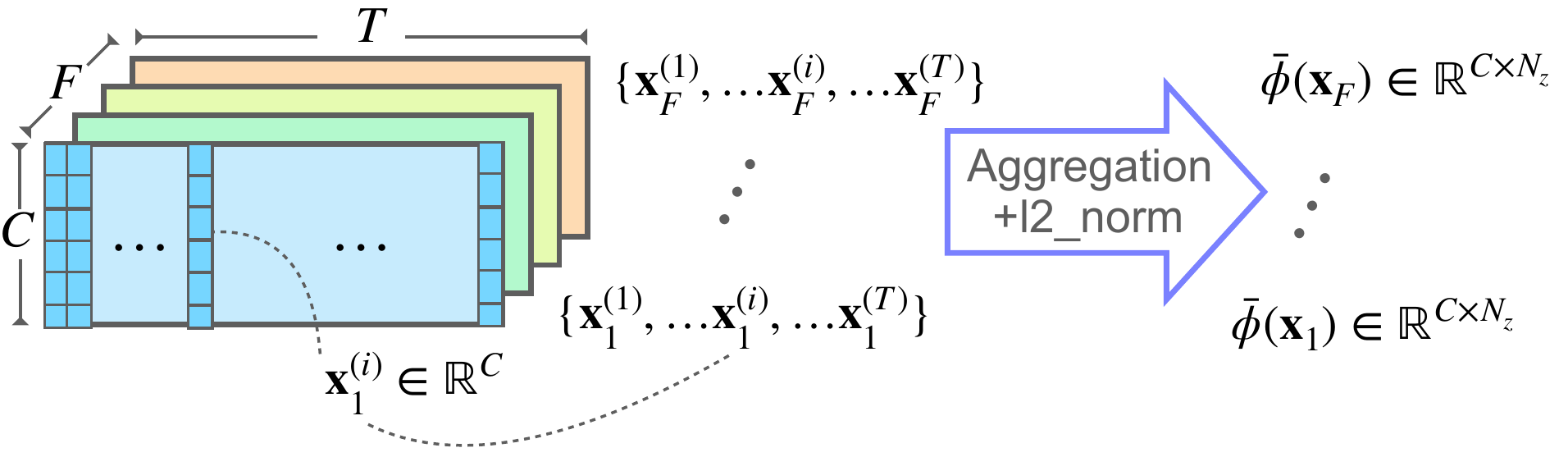}
  \caption{Illustration of organizing the input for transport-oriented feature aggregation for speaker embedding learning.}
  
  \label{fig:agg}
  \vspace{-1em}
\end{figure}

The pre-aggregation layer in a 2D-CNN speaker embedding network produces an output of dimension $C\times F\times T$, where $C$, $F$, $T$ are the number of channels, frequencies and time frames respectively. The size of the input set $\mathbf{x}$ is determined by $T$. The question is how to select the elements of $\mathbf{x}$. Our view is for speaker modeling, it is most appropriate to pick vectors along the channel axis as the feature representation, and keep the frequency and channel axis intact (i.e. no reshaping operation is applied to merge these two axes). This is because feature values along the frequency axis encode pertinent phonetic information, and their distribution reflects more about the speech content than speaker characteristics. Aggregation is therefore applied to $F$ individual sets of features, each corresponding to one frequency, as shown in Figure \ref{fig:agg}. Following the prior works  \cite{netvladaudio,lde,netvlad,deepten}, the aggregated vectors are $l_2$-normalized.

Incorporating the attention mechanism is another important consideration in speaker embedding learning. In the original scheme of trainable optimal transport embedding \cite{OTKE}, both the sample set and the reference set have uniform probability measure, i.e., $\mathbf{a}=\mathbf{1}/N_x$ and $\mathbf{b}=\mathbf{1}/N_z$ in Algorithm 1. While it is reasonable to assume no outlier in the reference set and hence assign all reference elements equal weight, this may not be appropriate for the sample set, especially when dealing with dynamically changing speech data. Intuitively features from certain time frames are more important than others in identifying speakers. In addition to learning a reference distribution $\mathbf{z}$ as stated in Section \ref{approach}, we propose to learn a reference element $\mathbf{u}\in\mathbb{R}^{d}$, and the weight assigned to the $i^{th}$ frame in Algorithm 1 is defined as
\begin{equation}\label{attention}
    \mathbf{a}=[a_1, \ldots,a_i, \ldots, a_{N_x}], \text{ }a_i=\frac{\exp{(\mathbf{u}^T\mathbf{x}^{(i)} )}}{\sum_{i=1}^{N_x}\exp{(\mathbf{u}^T\mathbf{x}^{(i)} )}}
\end{equation}

To summarize, given the pre-aggregation layer output of shape $C\times F \times T$, Algorithm 1 is applied to $F$ individual sets, each of which contains $T$ elements of dimension $C$, and that there are $F$ reference sets, as well as $F$ reference elements (if attention is to be implemented) to be learned.

\section{Experiments}
A toy example of classifying 1D distributions is described first to illustrate the merit of transport-oriented feature aggregation over statistics pooling. Then experiments on speaker verification are carried out to evaluate speaker embeddings learned with transport-oriented feature aggregation. We repeat all experiments 3 times with different random seeds and report the mean and standard deviation of our results.

In the following discussion, statistics pooling is denoted as \emph{Stats}, and transport-oriented feature aggregation is denoted as \emph{OT-r}, where \emph{OT} abbreviates \emph{optimal transport}, and \emph{r} is the size of the reference set.

\subsection{Toy example}
\begin{figure}[t]
  \centering
  \includegraphics[width=0.8\linewidth]{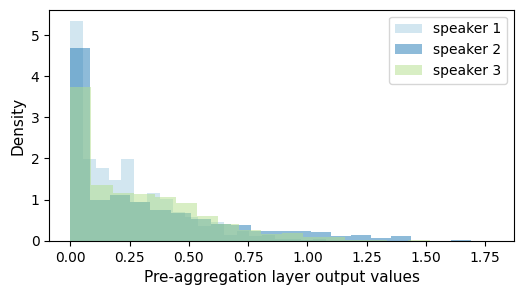}
  \caption{Histograms of the pre-aggregation layer outputs from a single channel in a ResNet-based speaker embedding network, trained and evaluated on Voxceleb2. (It is the first baseline in Section \ref{svexp}.) The overall structure of output value distribution is shared by most channels.}
  \label{fig:hist}
\end{figure}

In this example the goal is to learn class labels from i.i.d. samples drawn from different 1D distributions, each corresponding to a class. The experimental setup is developed from an empirical finding: the pre-aggregation layer output values are distributed in a similar pattern across most of the channels in a trained ResNet-based speaker embedding network, as illustrated in Figure \ref{fig:hist}. Their distributions are discontinuous at the value 0, presumably due to the use of ReLU activation. Based on this observation, we generate data from the following mixed Gamma distribution \cite{mixed_gamma}, which approximates the type of distributions observed in Figure \ref{fig:hist}.

\begin{equation}
  P(x)=\begin{cases}
    p, & \text{if $x=0$}.\\
    (1-p)\Gamma(k,\theta), & \text{if $x>0$}.
  \end{cases}
\end{equation}

We then consider $n=100$ distributions $\rho_1,...\rho_n$, each of which is a mixed Gamma, with the three parameters $p$, $k$ and $\theta$ picked uniformly on $[0.2,0.8]$, $[0.5, 2.5]$, $[0.2,1.0]$ respectively. A total of 10,000 samples per distribution are generated for training, and 1000 generated for testing. Here one sample refers to a set of 25 observations during training, and 50 when testing. The network is trained to identify the distribution on each given sample. 

\begin{figure}[t]
  \centering
  \includegraphics[width=1.0\linewidth]{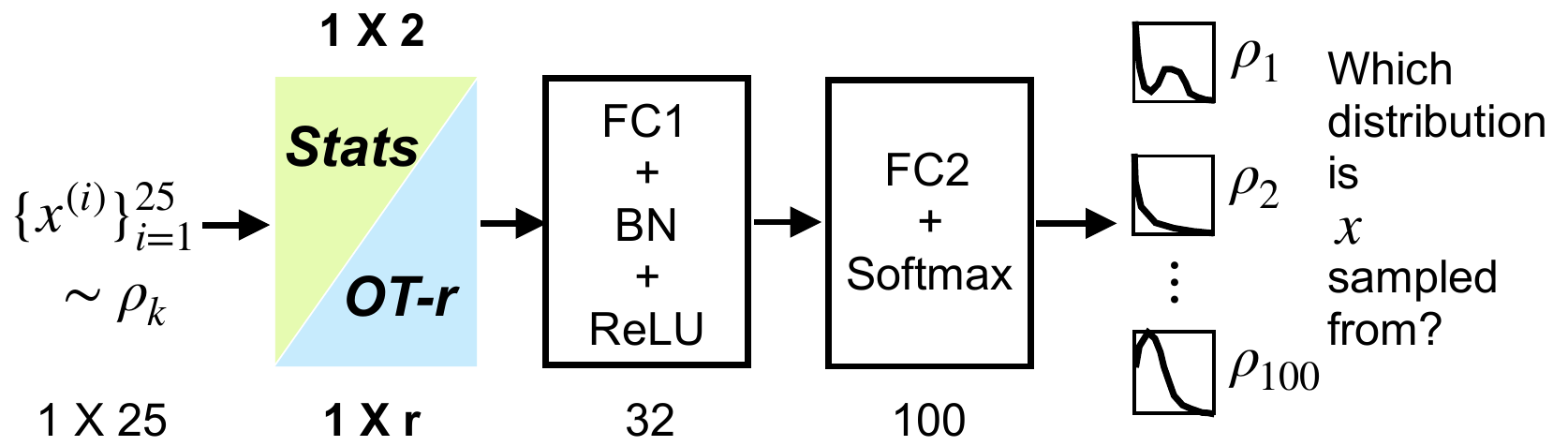}
  \caption{Illustration of the toy example, where numbers denote the layer output size, e.g. the shape of the input data is $1\times 25$. The batch dimension is omitted here.}
   \vspace{-1em}
  
  \label{fig:toy}
\end{figure}

We test two simple classifier networks. They have exactly the same architecture, except for their feature aggregation layers: \emph{Stats} v.s. \emph{OT-r}, as shown in Figure \ref{fig:toy}. Experimental results in Table \ref{tab:toy} suggest that the size of the reference set has a significant impact on the model performance for transport-oriented feature aggregation. It is noted that \emph{OT-2} performs significantly worse than \emph{Stats}. This is however expected because two observations are obviously too few to represent a distribution. As the reference size reaches 8, transport-oriented feature aggregation starts to substantially improve over the statistics baseline, suggesting that it is better at representing the distribution.

\begin{table}
		\caption{Experimental results on 1D distribution classification of the toy example.}
		\label{tab:toy}
		\centering
		\setlength\tabcolsep{5.5pt}
		\renewcommand{\arraystretch}{0.95}
 		\begin{tabular}{l|cccccc}
			\toprule
			Pooling & Stats & OT-2 & OT-4 & OT-8 & OT-16 & OT-32\\
			\midrule
			\midrule
			\multirow{2}{*}{Acc\%} & 20.6 & 6.8 & 21.6 & 26.4& 28.6 & 29.8 \\
			  & $\pm$0.4 & $\pm$0.9 & $\pm$1.2 & $\pm$1.4& $\pm$1.6 & $\pm$1.2\\
			\bottomrule
		\end{tabular}
 	\vspace{-0.2em}
	\end{table}

\subsection{Speaker verification task}\label{svexp}
\textbf{Data.} The development set of Voxceleb2 is used for training. It consists of about 1.1 M utterances from over $5,000$ speakers. No data augmentation or voice activity detection is applied. Performance evaluation is carried out with the three official Voxceleb1 test sets: cleaned Voxceleb1-O (\emph{original}), cleaned Voxceleb1-H (\emph{hard}), cleaned Voxceleb1-E (\emph{extend}).

\noindent\textbf{Network architecture.} The input features are 64-dimensional log Mel-Filterbank coefficients, computed with a window size of 25ms shifted by 10ms, and mean-normalized at utterance-level. The input data are first processed by a 2D convolutional layer and then fed to a ResNetSE-34 backbone. The strides and number of channels for each of the four blocks in the ResNet are $(1,2,2,2)$ and $(32,64,128,256)$ respectively. Squeeze and Excitation layers \cite{SE} are included in all blocks. An additional projection layer is added to reduce the ResNet output dimension from 256 to 64. The aggregation layer consists of 8 reference sets, plus 8 reference elements if attention is to be incorporated. The aggregated features are first $l_2$-normalized for each frequency group, then concatenated and compressed into a 256-dimensional speaker embedding. This embedding is fed to the final FC layer for classification. Details of the network architecture are given in Table \ref{tab:net}.

\noindent\textbf{Baselines.} Two baseline models are implemented. They have exactly the same network architecture as in Table \ref{tab:net} except for their aggregation layers: one using statistics pooling, and the other using multi-head self-attentive statistics pooling \cite{att2}. For the multi-head self-attention baseline, each head is responsible for one frequency group. The projection layer is not included in baseline models as there is no need to reduce dimension.

\begin{table}
		\caption{Architecture of the speaker embedding network. $F$, $T$, $N$ refer to the number of frequency bins, time frames and speakers respectively. $r$ is the size of the reference set. * indicates optionally used to incorporate attention. ($\cdot$) indicates the dimension where $l_2$-norm is applied on.}
		\label{tab:net}
		\centering
		\setlength\tabcolsep{3.5pt}
		\renewcommand{\arraystretch}{0.95}
 		\begin{tabular}{l l l}
			\toprule
			Layer & Structure & Output size\\
			\midrule
			\midrule
			Input & --- & $T\times F\times 1$\\[0.9ex]
			Conv2D & Conv$3\times3$, stride 1& $T\times F\times 32$\\[0.9ex]
			Backbone & ResNetSE34-layers & $T/8\times F/8 \times 256$\\[0.9ex]
			Projection & Conv$1\times1$, $256\Rightarrow 64$ & $T/8\times F/8 \times 64$\\[0.9ex]
			 & $F/8$ reference sets &\\
			{Aggregation} & $F/8$ reference samples* &  {$F/8\times(64\times r)$}\\
			&$l_2$-norm & \\[0.9ex]
			Reshape+FC & {$F/8\times64\times r \Rightarrow 256$} & 256\\[0.9ex]
			FC+Softmax & $256\Rightarrow N$ & $N$ \\
			\bottomrule
		\end{tabular}
			\vspace{-1.2em}
	\end{table}
	
\noindent\textbf{Training.} We use the Additive Angular Margin (AAM) loss \cite{aam} and set the scaling coefficient to 30. During training, we randomly crop a 2-second segment from each utterance and one batch contains 128 segments. All models are trained for 20 epochs using the Adam optimizer, with weight decay (5e-5), and a step decaying learning schedule (initialized to 1e-3 and divided by 10 every 8 epochs). This learning schedule is optimized on the two baselines and we keep it unchanged for training transport-based models. For transport-based models, the maximum number of iterations in Algorithm 1 is set to 20, and we use $\epsilon=1.0$ for the entropy term.

\noindent\textbf{Evaluation.} At test time, we sample 4-second cropped segments from each utterance in the trial with a shift of 1.0 second. We compute the cosine distance between every possible pair of speech segments and use the averaged cosine distance as the score for that trial. No score normalization is applied.

\noindent\textbf{Results.} Table \ref{tab:eer} shows that with a reference set of size 16 (\emph{OT-16}), transport-oriented aggregation achieves comparative results to the stronger multi-head self-attention baseline. We can also see that a larger reference set generally leads to better performance for transport-based models, but the benefits start to tail off as the size of the reference set reaches 16. Similar patterns are also observed in the toy example. As we increase the size of the reference set to 32, and include the attention as described in equation (\ref{attention}), the corresponding transport-based model (\emph{OT-32+Att}) gives the best results among all models.

\noindent\textbf{Computational cost.} Though the Sinkhorn iteration in Algorithm 1 is not the main computational bottleneck and usually converges within 10 iterations in our experiments, a relative drop of 12\% in model's inference speed is observed. 
\begin{table}
		\caption{EER\% test results on the Original, Hard, and Extended VoxCeleb test sets. Cosine distance is used as the backend.}
		\label{tab:eer}
		\centering
		\setlength\tabcolsep{4.5pt}
 		\begin{tabular}{l|ccc}
			\toprule
			Aggregation & Voxceleb1-O & Voxceleb1-H & Voxceleb1-E\\
			\midrule
			\midrule
			Stats & 1.44$\pm$0.04 & 2.67$\pm$0.02 & 1.50$\pm$0.02\\
			Stats+Att& 1.38$\pm$0.01 & 2.51$\pm$0.02& 1.45$\pm$0.01\\
			\midrule
			OT-8 & 1.37$\pm$0.04 & 2.60$\pm$0.01 & 1.50$\pm$0.02\\
			OT-16 & 1.34$\pm$0.04 & 2.52$\pm$0.01 & 1.48$\pm$0.02\\
			OT-32 & 1.37$\pm$0.01 & 2.49$\pm$0.02 & 1.43$\pm$0.02\\
			OT-32+Att & \textbf{1.32$\pm$0.03} & \textbf{2.42$\pm$0.02} & \textbf{1.42$\pm$0.01}\\
			\bottomrule
		\end{tabular}
		
		\vspace{-1.2em}
		
	\end{table}

\section{Conclusion and future work}
We described the idea of distinguishing speakers by measuring the optimal transport distance between distributions of extracted features, which aims at capturing pertinent speaker information that are not well represented by individual statistical measures. This can be achieved by using transport-oriented feature aggregation. Experiments on speaker verification have shown improvement over the standard statistics pooling and its attentive variant. A 12\% relative drop in inference speed has been observed and attributed to the computation required for solving optimal transport plans. We plan to explore the use of optimal transport distance as a regularization loss in future work. This is in line with the proposed idea of distinguishing speakers based on feature distributions, but would not affect the model architecture and inference speed.

\section{Acknowledgements}
Yusheng Tian and Jingyu Li are supported by the Hong Kong Ph.D. Fellowship Scheme of the Hong Kong Research Grants Council.

\bibliographystyle{IEEEtran}

\bibliography{mybib}


\end{document}